\documentclass[twocolumn, showpacs, preprintnumbers, amsmath, amssymb]{revtex4}

\usepackage{graphicx}
\usepackage{dcolumn}
\usepackage{bm}

\begin{document}

\title{Exactly soluble spin-$\frac{1}2$ models on three-dimensional
lattices and non-abelian statistics of closed string excitations}

\author{Tieyan Si}\author{Yue Yu}
\affiliation{Institute of Theoretical Physics, Chinese Academy of
Sciences, P.O. Box 2735, Beijing 100080, China}

\date{\today}

\begin{abstract}
Exactly soluble spin-$\frac{1}2$ models on three-dimensional
lattices are proposed by generalizing Kitaev model on honeycomb
lattice to three dimensions with proper periodic boundary
conditions. The simplest example is spins on a diamond lattice
which is exactly soluble. The ground state sector of the model may
be mapped into a $p$-wave paired state on cubic lattice. We
observe for the first time a topological phase transition from a
gapless phase to a gapped phase in an exactly soluble spin model.
Furthermore, the gapless phase can not be gapped by a perturbation
breaking the time reversal symmetry. Unknotted and unlinked Wilson
loops arise as eigen excitations, which may evolute into linked
and knotted loop excitations. We show that these closed string
excitations obey abelian statistics in the gapped phase and
non-abelian statistics in the gapless phase.

\end{abstract}
\pacs{75.10.Jm,03.67.Pp,71.10.Pm}

\maketitle

\noindent {\it Introductions:} Non-abelian anyons in two
dimensions provide a promising candidate for  quantum computation.
which is topologically protected from decoherence \cite{freedman}.
Recently, this topic has attracted great interests \cite{das}. The
most possible system in which the quasiparticles obey non-abelian
statistics is two-dimensional electron gas in a fractional quantum
Hall state with $\nu=5/2$ \cite{read, xia}.

In search for non-abelian statistics in exact soluble models,
Kitaev proposed an exactly soluble spin model on honeycomb lattice
and showed that in a wider parameter region, the vortex
excitations obey non-abelian statistics \cite{kitaev}. Since the
exotic statistics of the vortices, Kitaev model has attracted many
research interests \cite{km,chn}. It has been shown that Kitaev
model is in the same universality class with $p_x+ip_y$-wave
paired state \cite{yw}.

The point-like anyons are strictly restricted in two dimensions.
In three dimensions (3-d), the point-like particles can only be
either bosons or fermions. However, the exotic statistics may
arise in the closed string excitations.   In a seminal paper
\cite{witten}, Witten has shown that the Wilson loops in 3-d
 Chern-Simons field theory obey non-abelian statistics, which
is closely related to the monodromy matrix in conformal field
theory \cite{ms}. The statistics of unknotted, unlinked closed
strings is described by loop braid groups\cite{baez}. The quantum
loop gas applied to topological quantum computation is a rapidly
developing field \cite{loops}. Topological quantum order in brane
systems was also an interesting subject \cite{brane}.

In this paper, we will generalize Kitaev model to that on a
diamond lattice as well as multi-layer honeycomb lattices in which
the closed string excitations obey non-abelian statistics. This
3-d generalization of Kitaev model is also exactly soluble and the
ground state sector is equivalent to 3-d $p$-wave paired  state.
The phase diagram consists of two topological phases, a gapped one
and gapless one. The gapped phase is a strong pairing phase whose
topological nature is characterized by Hopf invariant and then the
ground state is topologically trivial. The singularity of the Hopf
mapping in the gapless phase implies that the ground state is
topologically non-trivial. Removing the singular points, a
non-zero winding number may appear. Thus, in the gapped phase, the
closed strings obey abelian anyonic statistics while in the
gapless phase, the strings obey non-abelian statistics, which is
tantamount to Wilson loops in SU(2)$_2$ Chern-Simons field theory
\cite{witten}. The Majorana fermion excitations in the gapless
phase is always gapless even there is a perturbation with
time-reversal symmetry breaking. A topological phase transition
between gapped and gapless phases \cite{vol} is found for the
first time in an exactly soluble model.

\begin{figure}
\begin{center}
\includegraphics[width=0.35\textwidth]{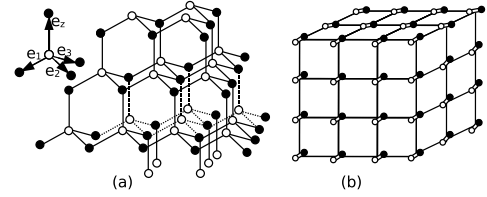}
\caption{\label{fig:Fig. 1} The diamond and cubic lattices. (a)
The diamond lattice and four types of links. ${\bf e}_i$ (i=1,2,3)
corresponding to $x$-, $y$-, and $z$-links and ${\bf e}_z$ to
$v$-links. (b) The equivalent cubic lattice.}
\end{center}
\vspace{-0.8cm}
\end{figure}

 \noindent {\it Model and
Solution:} We begin with a Kitaev-type coupled spin-$\frac{1}2$
model on a diamond lattice (Fig. \ref{fig:Fig. 1}a), whose
Hamiltonian reads
\begin{eqnarray}
H_d&=&\sum_{\textbf{i}}(J_{x}\sigma^{x}_{\textbf{i}}\sigma^{x}_{\textbf{i}+\textbf{e}_{1}}
+J_{y}\sigma^{y}_{\textbf{i}}\sigma^{y}_{\textbf{i}+\textbf{e}_{2}}
\nonumber\\
&+&J_{z}\sigma^{z}_{\textbf{i}}\sigma^{z}_{\textbf{i}+\textbf{e}_{3}}+
J_{v}\sigma^{z}_{\textbf{i}}\sigma^{z}_{\textbf{i}+\textbf{e}_{z}}),
\end{eqnarray}
where the translational invariant vector is
$\textbf{i}=m\textbf{e}_{1}+n\textbf{e}_{2}+l\textbf{e}_{3}-(m+n+l)\textbf{e}_{z}$.
 This Hamiltonian
 has a set of integrals of motion
$\{S_{p},p=1,2,...\}$ in which anyone commutes with the
Hamiltonian as well as another member in the set. $S_{p}$ is a
loop operator, i.e., an unknotted and unlinked closed string
operator along $\textbf{e}_{\perp3}=\textbf{e}_{2}-\textbf{e}_{1}$
if we take periodic boundary condition along
$\textbf{e}_{\perp3}$. The explicit form of $S_{p}$ is
$S_{p}=\prod_{\textbf{i}=m\textbf{e}_{\perp3}}
\sigma^{x}_{\textbf{i}+\textbf{e}_{1}}\sigma^{x}_{\textbf{i}}
\sigma^{y}_{\textbf{i}}\sigma^{y}_{\textbf{i}+\textbf{e}_{2}}$
which reduces to $S_{p}=\prod_{m=0}^{n-1}i^{n}
\sigma^{z}_{m\textbf{e}_{\perp3}}$ under the periodic constrain
$\vec{\sigma}_{\textbf{i}}=\vec{\sigma}_{\textbf{i}+n\textbf{e}_{\perp3}}$.
Each $S_{p}$ projects a sector of the total eigenspace of the
Hamiltonian $H_d$ since $[S_{p},H_d]=0$. This guarantees the
existence of exact solutions.

Expressing spins by Majorana fermions \cite{kitaev}, one has
$\sigma^{x}=ib_{x} c, \sigma^{y}=ib_{y}c$, and
$\sigma^{z}=ib_{z}c$ with constraint $b_xb_yb_zc=1$. Defining
$u_{ij}=ib_i^{x,y}b_j^{x,y}$ for the $x$- and $y$-links and
$u_{ij}=ib_i^zb_j^z$ for $z$- and $v$-links, the Hamiltonian $H_d$
reads
\begin{eqnarray}
H_d= &=&iJ_z\sum_{s} u_{s,bw}c_{sb}c_{sw}\nonumber\\&
+&iJ_x\sum_su_{s,bw}(c_{sb}c_{s-{\bf e}_1,w}-c_{s,w}c_{s-{\bf
e}_1,b})\nonumber\\&+&i
J_x\sum_s u_{s,bw}(c_{s,b}c_{s-{\bf e}_1,w}+c_{s,w}c_{s-{\bf e}_1,b}) \nonumber\\
&+&J_y~\& ~J_v~{\rm partners}\label{gh},
\end{eqnarray}
where $s$ is the position of the $z$-links. Note that
$[H_d,u_{ij}]=0$, $u_{ij}^2=1$ and $S_p$ is the product of a set
of $u_{ij}$. Therefore, one may take $u_{ij}=\pm 1$ and the eigen
value of $S_p$ may be determined by product of $u_{ij}$.
 According to Lieb's
theorem \cite{lieb}, the ground state is included in the sector
with all $u_{s,bw}=1$, which is consistent with all $S_P=1$. If
there are odd numbers of $u_{bw}=-1$ along a loop, this gives
$S_P=-1$.  This is a loop excitation and in fact is a nontrivial
Wilson loop. Defining the link fermions living in $z$-links,
$d_s=(c_{s,b}+ic_{s,w})/2, ~~~d^\dag_s=(c_{s,b}-ic_{s,w})/2$ and
deforming the lattice to a cubic lattice (Fig. \ref{fig:Fig. 1}b)
in a similar way deforming the honeycomb to a square lattice in
two dimensions \cite{chn,yw}, the loop excitation-free Hamiltonian
may be written as
\begin{eqnarray}
H_0&=&\sum_{\bf p} \xi_{\bf p}d^\dag_{\bf p}d_{\bf
p}+\frac{\Delta_{1,\bf p}}2(d_{\bf p}^\dag d^\dag_{-{\bf p}}+d_{\bf p} d_{-{\bf p}})\nonumber\\
&+&i\frac{\Delta_{2,\bf p}}2(d_{\bf p}^\dag d^\dag_{-{\bf
p}}-d_{\bf p} d_{-{\bf p}})
\end{eqnarray}
where $d_{\bf p}$ is the Fourier component of $d_s$; the
dispersion and the pairing functions are
\begin{eqnarray}
&&\xi_{\bf p}=J_z-J_x\cos p_x-J_y\cos
p_y-J_v\cos p_z,\\
&&\Delta_{a,\bf p}=\Delta_{ax}\sin p_x+\Delta_{ay}\sin
p_y+\Delta_{az}\sin p_z,\quad a=1,2.\nonumber
\end{eqnarray}
This is a $p$-wave paired state. At present, $\Delta_{1b}=0$ and
$\Delta_{2b}=J_b$ with $b=x,y,v$. Furthermore, one can follow a
previous work by Wang and one of the authors (Y.Y) \cite{yw} and
introduce three-spin couplings and four-spin couplings etc. This
leads to the parameters $J_{x,y}$ and $\Delta_{a;x,y}$ become
tunable. To let $J_z$ and $\Delta_{az}$ be tunable, one needs to
add terms like $\sigma^z_{s,b}\sigma^z_{s+{\bf
e}_z,b}+\sigma^z_{s,w}\sigma^z_{s+{\bf e}_z,w}$ and
$\sigma^z_{s,b}\sigma^z_{s+{\bf e}_z,w}$. These are not nearest
neighbor couplings. One can rewritten them as
$\sigma^z_{s,b}(b_{s+{\bf e}_z,w}^zb_{s+{\bf
e}_z,w}^z)\sigma^z_{s+{\bf
e}_z,b}+\sigma^z_{s,w}(b_{s,b}^zb_{s,b}^z)\sigma^z_{s+{\bf
e}_z,w}$ and $\sigma^z_{s,w} (b_{s,b}^zb_{s,b}^z)(b_{s+{\bf
e}_z,w}^zb_{s+{\bf e}_z,w}^z)\sigma^z_{s+{\bf e}_z,b}$ because
$b_zb_z=1$. Thus, these terms can also be bilinear since $u_{ij}$
are correctly inserted. Thus, the model with these terms added is
still exactly soluble but these pairing parameters $\Delta_{ab}$
become tunable. Hereafter, we discuss this general $p$-wave paired
state. The quasiparticle excitations are governed by the BdG
equations
\begin{eqnarray}
E_{\bf p}u_{\bf p}=\xi_{\bf p}u_{\bf p}-\Delta^*_{\bf p}v_{\bf
p},~~E_{\bf p}v_{\bf p}=-\xi_{\bf p}v_{\bf p}-\Delta_{\bf p}u_{\bf
p} \label{bdg}
\end{eqnarray}
where $E_{\bf p}=\sqrt{\xi_{\bf p}^2+(\Delta_{1,\bf
p})^2+(\Delta_{2,\bf p})^2}$ is the quasiparticle dispersion,
$\Delta_{\bf p}=\Delta_{1,{\bf p}}+i\Delta_{2,{\bf p}}$, and
$(u_p,v_p)$ are the coherence factors with $|u_{\bf
p}|^2=\frac{1}2(1+\frac{\xi_{\bf p}}{E_{\bf p}}),|v_{\bf
p}|^2=\frac{1}2(1-\frac{\xi_{\bf p}}{E_{\bf p}})$ and $v_{\bf
p}/u_{\bf p}=-(E_{\bf p}-\xi_{\bf p})/\Delta^*_{\bf p}$.

The phase diagram can be drawn in a similar way to that for the
two-dimensional copy \cite{yw}. The phase boundary is determined
by $\xi_{\bf p}=0$ for any $\Delta_{ab}$. That is, the band
insulator/free Fermi gas transition determines the phase boundary,
which is given by $|\cos p_a^*|=1$ with ${\bf p}^*=(0,0,0)$,
$(0,0,\pm \pi)$,$(0,\pm \pi,0)$,$(\pm \pi,0,0)$, $(0,\pm \pi,\pm
\pi)$, $(\pm \pi,0,\pm \pi)$,$(\pm \pi,\pm \pi,0)$ and $(\pm
\pi,\pm \pi,\pm \pi)$ where $J_z\pm J_x\pm J_y\pm J_v=0$.  The
phase where $\xi_{\bf p}>0$ is always gapped. In the $p$-wave
sense, this is the strong pairing phase as that in two dimensions
\cite{rg}. However, out of the gapped phase, there are a pair of
${\bf p^*}$ so that $E_{\bf p^*}=0$ for general $\Delta_{ab}$.
This means that there is a phase transition from the gapped phase
to a gapless phase even the time reversal symmetry (or the
generalized inversion symmetry \cite{yw}) is broken.

 \noindent{\it Topology of Phases:} Now, the
question is that whether the phase transition from the gapped  to
gapless phases is a topological phase transition. First, we do not
observe the spontaneous breaking of any continuous symmetry. This
implies the phase transition may be topological. However, in
general, a topological phase transition requires an energy gap
between the degenerate ground state and any excitation state. To
check if the phase with a gapless Majorana fermion excitation is
topologically non-trivial, we calculate the topological invariants
of both phases in the continuum limit. The momentum space is
$D^2\times S^1$ due to the periodic boundary condition. Note that
the coherent function $\psi^\dag=(u^*,v^*)$ defines a mapping from
${\bf p}\in D^2\times S^1$ to $(u,v)\in S^2$. The unit vector
${\bf m}=\psi^\dag\vec \sigma \psi=({\rm Re}\Delta_{\bf p},-{\rm
Im}\Delta_{\bf p}, \xi_{\bf p})/E_{\bf p}$ parameterizes this
mapping. If we compact $D^2\times S^1$ to $S^3$ by defining ${\bf
m=m}_0$ on the boundary torus $T^2$, the mapping is called the
Hopf mapping. Associated with Hopf mapping, there is a topological
invariant, the Hopf invariant, which is defined by
\begin{eqnarray}
{\rm Hf}=\frac{1}{8\pi^2}\int _{S^3}d^3p~\epsilon_{ijk}A^i
F^{jk}.\label{hf}
\end{eqnarray}
with $A_i(p)=\frac{i}2(\psi^\dag\partial_{p_i}
\psi-\psi\partial_{p_i} \psi^\dag)$ and $F_{ij}(p)=\partial_{p_i}
A_j(p)-\partial_{p_j} A_i(p)= {\bf m}\cdot (\partial_{p_i} {\bf
m}\times
\partial_{p_j} {\bf m})$. This is an abelian Chern-Simons in the
momentum space. On the other hand, we know that the Hopf invariant
describes the linking numbers of closed strings \cite{pol}. Due to
the duality between the cubic lattice and dual momentum lattice,
the linking number in momentum space is equal to that in the
co-ordinate space. For the strong pairing ground state, this
linking number is zero because there is no closed string
excitation. One may also directly prove that
$(u(\infty),v(\infty))=(u(0),v(0))=(0,1)$ in the ground state and
thus, Hf=0.

Note that the unit vector ${\bf m(p^*)}$ is singular in the
gapless phase because at Dirac points ${\bf p}^*$, $E_{\bf
p^*}=0$. For a set of given parameters in the gapless phase, there
are two such Dirac points except at the phase boundary where ${\bf
p}^*_1={\bf p}^*_2=0$. The unit vector ${\bf m}$ can not be
defined in these singular points. To see the topological property,
one may pick off these two Dirac points for given parameters.
Then, $S^3$ is reduced to $S^3-\{{\bf p}^*_1,{\bf p}^*_2\}$, which
can contract to an $S^2$ while at the phase boundary it is $D^3$
which can contract to origin. The later fact means there is a
discontinuity of the mapping at the phase boundary. Inside of the
gapless phase, the Hopf mapping now is reduced to a mapping
$(u,v)$ from $S^2$ to $S^2$. Taking $(p_x,p_y)$ as the coordinate
in the source $S^2$, then the winding number of this mapping
\cite{volovik} is the same as that defined in two-dimensional
Kitaev-type model
\begin{eqnarray}
\nu&=&\frac{1}{4\pi}\int _{S^2}dp_xdp_y (\partial_{p_x}{\bf
m(p)})\times (\partial_{p_y}{\bf m(p}))\cdot {\bf m(p)}
\end{eqnarray}
Since $(u(\infty),v(\infty))=(1,0)$ is the north pole of the
target $S^2$ and $(u(0),v(0))=(0,1)$ is the south pole, we know
that $\nu=1$. Therefore, the ground state of the gapless phase is
topologically non-trivial \cite{nick}. ( As explained in
\cite{nick}, although there is a phase transition from gapless to
gapped phase transition in Kitaev's original model, both phases
are topologically trivial.)

\begin{figure}
\begin{center}\label{multilayer}
\includegraphics[width=0.32\textwidth]{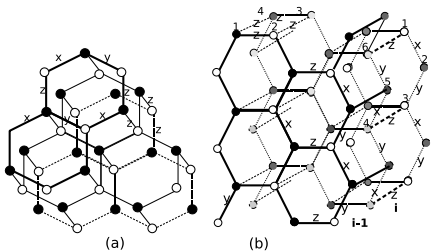}
\caption{(a) The bonding configuration of the multilayer model
$H_{z}$. (b) The bonding configuration of another multilayer model
mentioned in \cite{E12}.}
\end{center}
\vspace{-0.9cm}
\end{figure}

\noindent{\it Loops, Knots and Links:} We now define the loops,
knots and links. Loops are excited if $S_P=-1$. Two loops may
merge into a single loop by operation of plaquette operator. Knot
and link excitations are created by this operations. To see these
operators more clearly, we show it by using a model on a
multi-layer honeycomb lattice which is in the same university
class with the diamond lattice model in the long wave length
limit, i.e., they both are described by a 3-d $p$-wave paired
state. The links arrangement of the new model is shown in Fig. 2a
and the Hamiltonian reads
\begin{eqnarray}\label{Hami-diamond1}
H_z&=&\sum_{\textbf{i}}J_{x}(\sigma^{x}_{\textbf{i}}\sigma^{x}_{\textbf{i}+\textbf{e}_{1}}
+\sigma^{x}_{\textbf{i}-\textbf{e}_{z}}\sigma^{x}_{\textbf{i}-\textbf{e}_{z}+\textbf{e}_{2}})\nonumber\\
&+&J_{y}(\sigma^{y}_{\textbf{i}}\sigma^{y}_{\textbf{i}+\textbf{e}_{2}}
+\sigma^{y}_{\textbf{i}-\textbf{e}_{z}}
\sigma^{y}_{\textbf{i}-\textbf{e}_{z}+\textbf{e}_{1}})\nonumber\\
&+&J_{z}(\sigma^{z}_{\textbf{i}-\textbf{e}_{3}}\sigma^{z}_{\textbf{i}}
+\sigma^{z}_{\textbf{i}-\textbf{e}_{z}-\textbf{e}_{3}}\sigma^{z}_{\textbf{i}-\textbf{e}_{z}}),\nonumber\\
&+&J_{v}(\sigma^{z}_{\textbf{i}-\textbf{e}_{z}}\sigma^{z}_{\textbf{i}}
+\sigma^{z}_{\textbf{i}-\textbf{e}_{3}}\sigma^{z}_{\textbf{i}-\textbf{e}_{3}+\textbf{e}_{z}}),
\end{eqnarray}
where the fundamental translational invariant vector is
$\textbf{i}=m(\textbf{e}_{1}-\textbf{e}_{3})+n(\textbf{e}_{2}-\textbf{e}_{3})+2l\textbf{e}_{z}$,
with basis $\textbf{e}_{1}=(\frac{1}{2},\frac{\sqrt{3}}{2},0)$,
$\textbf{e}_{2}=(\frac{1}{2},-\frac{\sqrt{3}}{2},0)$,
$\textbf{e}_{3}=(-1,0,0)$, $\textbf{e}_{z}=(0,0,1)$. Again, there
exist a series of string operators $S_{p}=\prod_{m=0}^{n-1}i^{n}
\sigma^{z}_{m\textbf{e}_{\perp3}}$ which commute with the
Hamiltonian under periodic boundary condition. Using Majorana
fermions, we obtain bilinear fermionic Hamiltonian and the closed
string excitation-free Hamiltonian has a dispersion
$E=|E_{1}E_{2}|^{1/2}$\cite{E12}. The gapless phase falls in the
area given by the inequalities
$|J_{v}|\leq|J_{x}|+|J_{y}|+|J_{z}|$,
$2|J_{i}J_{j}|\leq|J_{v}|^{2}+(|J_{x}|+|J_{y}|+|J_{z}|)^{2}$
$(i,j=x,y,z)$. One may prove that $H_z$ is also equivalent to a
$p$-wave paired model.

The lattice for $H_z$ is topologically equivalent to a solid torus
$D^{2}\times{S^{1}}$ with $\textbf{e}_{\perp3}$ as its tangent
vector(Fig. 3a). All the loop operators following the geodesic
circles commute with the Hamiltonian and they are good quantum
numbers. Two neighboring loops following zigzag geodesic circle
are the product of plaquette operators $\hat P$ sandwiched between
them, $\hat{P}_{1}\hat{P}_{2}\cdots
\hat{P}_{n-1}=S^{\textbf{i}}_{p}{S^{\textbf{i}+\textbf{e}_{3}}_{p}}$,
where
$\hat{P}_{j}=\sigma^{x}_{j}\sigma^{y}_{2j}\sigma^{z}_{3j}\sigma^{x}_{4j}\sigma^{y}_{5j}\sigma^{z}_{6j}$
is the plaquette operator defined on the hexagon. In this way, we
merge two loops into a single loop. These loops form a loop gas.
Links and knots can be obtained by twitted boundary conditions
(See Figs. 3b and 3c). These knots and links may also be thought
$Z_2$ Wilson loops but they are not exact eigen excitations of
$H_z$. However, the non-commutativity between $H_z$ and the knots
and links may only happen in the bonds connecting two geodesic
circles. Therefore, one can ignore this non-commutativity in the
thermodynamic and long wave length limits and takes these Wilson
loops to be quasi-exact eigen excitations. The above analysis may
also be done on diamond lattice. However, since there is a shift
between two adjacent layers, the illustrations are not so direct.

\begin{figure}
\begin{center}
\includegraphics[width=0.5\textwidth]{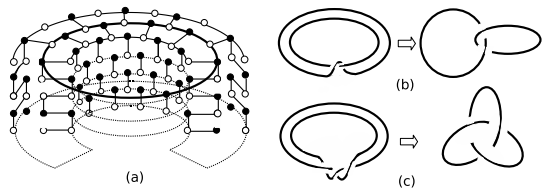}
\caption{(a) The topological equivalent manifold of model $H_{z}$,
a solid torus. (b) A typical link excitation (Hopf link). (c) A
knot excitation (trefoil).}
\end{center}
\vspace{-0.8cm}
\end{figure}

\noindent{\it Statistics:}
 One can calculate the linking number of a closed string
configuration. In the gapped phase, the Hopf invariant (\ref{hf})
counts the linking number of the dual links in momentum space,
which is equal to the linking number on the original co-ordinate
space. The fact that the Hopf invariant is the abelian
Chern-Sinoms implies these closed string excitations obey abelian
statistics.

\begin{figure}
\begin{center}
\includegraphics[width=0.38\textwidth]{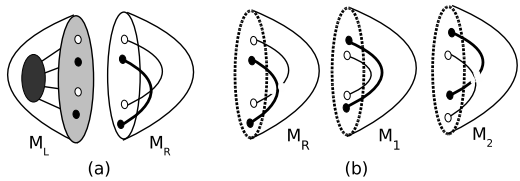}
\caption{(a) The left and right pieces of $S^3$ cut by $S^2$. (b)
Three states $L_+, L_0$ and $L_-$ of the right piece, which are
related by the matrix $B$.}
\end{center}
\vspace{-0.5cm}
\end{figure}

To see the closed string excitations obey non-abelian statistics
in the gapless phase, we follow Witten's discussions in dealing
with knots and links in SU(N)$_k$ Chern-Simons field theory
\cite{witten}. We compact the 3-d lattice to $S^3$ in the
continuum limit and then the Wilson loops are almost exact eigen
excitations. We consider two Wilson loops in $S^3$. Using an $S^2$
cuts $S^3$ into two pieces as shown in Fig. 4a, the left piece
$M_L$ may contain very complicated stuff but the right piece $M_R$
is simple. Then, two Wilson loops puncture four points which are
corresponding to creating two spin fields $\sigma$ and destroying
two $\sigma$ on $S^2$. The low energy limit of BdG equations
(\ref{bdg}) is the 3+1-dimensional Dirac equations. Hence, the
Hilbert space of the massless Majorana fermions restricted on
$S^2$ is determined by $c=1/2$ conformal field theory. The fusion
rule $\sigma\cdot \sigma=1+\psi$ implies the Hilbert space on this
punctured $S^2$ is two-dimensional. Therefore, three states
$L_+,L_0$ and $L_-$ showed in Fig. 4b are linear-dependent. By
using the monodromy $B$ matrix \cite{ms}, these states may relate
to each other through
\begin{eqnarray}
L_0=BL_+,~~~L_-=BL_0=B^2L_+,
\end{eqnarray}
where the $B$ matrix is not proportional to identity. Obviously,
the operation through the $B$ matrix is an exchange of these
vortices $\sigma$ and then they obey non-abelian statistics. Any
spin-$\frac{1}2$ has an SU(2) gauge symmetry if the spin is
represented by fermion or boson operators because the Hilbert
space is enlarged \cite{aff,yu}. So is the present model. Fixing
$u_{ij}=\pm 1$ is corresponding to a gauge fixing, i.e., reducing
the gauge symmetry from SU(2) to $Z_2$. After gauge
transformation, $\sigma$ carrys a unit SU(2) charge. By using the
non-ablelian bosonization \cite{wib}, the massless Majorana
fermion theory on $S^2$ is bosonized to an SU(2)$_2$/U(1) coset
Wess-Zumino-Witten model. The $k=2$ turns out that the Wilson
loops indeed obey non-abelian statistics in the sense Witten
defined. Due to the gauge group is in its defining representation,
the $B$ matrix has two eigenvalues
$\lambda_1=e^{-i\pi/8},~\lambda_2=-e^{i3\pi/8}$. Thus, the three
states in Fig. 4b satisfy the {\it skein relation}
\begin{eqnarray}
-qL_++(q^{1/2}-q^{-1/2})L_0+q^{-1}L_-=0,
\end{eqnarray}
with $q=e^{i\pi/2}$. Witten took this relation as the definition
of knot polynomials on $S^3$ \cite{witten}. As Witten pointed out,
the physical meaning of the skein relation is clear: all links and
knots can be resolved by the skein relation. Therefore, if we know
the physical behavior of unknotted and unlinked quantum loop gas,
the system with knots and links may be understood. This means that
the study of the quantum loop gas is fundamental \cite{loops}.

In conclusions, we constructed a 3-d exactly soluble spin model
with closed string excitations obeying non-abelian statistics.
This is closely related to Witten's original proposal to the
statistics of the Wilson loops in the Chern-Simons field theory. A
topological phase transition between gapless and gapped phases
 was first time predicted.

The authors thank Nick Read for useful discussions. They benefited
from participating the Program on Quantum Phases of Matter in
Kavli Institute for Theoretical Physics China. This work was
supported in part by the National Natural Science Foundation of
China.


\begin{thebibliography}{99}
\bibitem{E12} The dispersion is given by $E_{1}=e^{2iq_{z}}J^{2}_{v}-e^{i(q_{x}+2q_{y})}(J^{2}_{x}+J^{2}_{y})
-e^{i(q_{x}+(2-\sqrt{3})q_{y})}(1+e^{2i\sqrt{3}}q_{y})J_{x}J_{y}
-e^{\frac{1}{2}i(q_{x}+(2-\sqrt{3})q_{y})}(J_{x}J_{z}-J_{y}J_{z})
+e^{\frac{1}{2}i(q_{x}+(2+\sqrt{3})q_{y})}(J_{x}J_{z}-J_{y}J_{z})
+J^{2}_{z}$,
$E_{2}=e^{i(q_{x}+4q_{y})}J^{2}_{v}-e^{i(5-\frac{3\sqrt{3}}{2})q_{y}+i2q_{z}}
\{-e^{\frac{3}{2}i(\sqrt{3}-2)q_{y}}(J^{2}_{x}+J^{2}_{y})
-e^{\frac{1}{2}i[2q_{x}+(\sqrt{3}-2)q_{y}]}J^{2}_{z}
-(e^{\frac{1}{2}i(\sqrt{3}-6)q_{y}}+e^{\frac{1}{2}i(5\sqrt{3}-6)q_{y}})J_{x}J_{y}
-e^{\frac{1}{2}i(q_{x}+2(\sqrt{3}-2)q_{y})}(e^{i\sqrt{3}q_{y}}-1)(J_{x}-J_{y})J_{z}\}.
$ One may construct another exactly solube model which is on the
lattice as shown in Fig. 2b. However, the dispersion is more
complicated.
\end{thebibliography}
\end{document}